\documentclass{IET-Conf-Paper}

\usepackage{tikz}
\usetikzlibrary{arrows.meta,positioning}

\begin{document}

\title{GRID-FORMING CONTROL OF OFFSHORE WIND THROUGH A DIODE-RECTIFIER HVDC SYSTEM: A 2~GW CASE STUDY}

\author{Ishi Nigam\ad{1}\corr, Saif Alsarayreh\ad{1}, Tuanku Badzlin Hashfi\ad{1}, Milo\v{s} Cvetkovi\'c\ad{1}, Joris van Doorn\ad{2}, Aleksandra Leki\'c\ad{1}}

\address{\add{1}{Department of Electrical Sustainable Energy, Delft University of Technology, Delft, The Netherlands}
\add{2}{TenneT TSO B.V., Arnhem, The Netherlands}
\email{ishinigam@tudelft.nl}}


\begin{abstract}
\label{0_abstract}
Diode-rectifier-unit (DRU)-based high-voltage direct current (HVDC) transmission considerably reduces the volume and weight of the offshore converter platform for remote wind farms and, hence, its cost, but shifts the task of forming and regulating the offshore AC network to the wind-turbine converters. This paper examines which grid-forming control strategy is the most suitable for DRU power transfer. The DRU commutation relationship shows that the delivered active power depends on the offshore voltage magnitude (but not on the converter voltage angle), so the angle-based active-power loop of virtual synchronous machine control is not closed through the rectifier. A P-V/Q-$\omega$ grid-forming architecture, which regulates active power via the voltage magnitude and derives frequency from the reactive power error, is therefore adopted. Electromagnetic transient simulations of a 1~GW aggregated grid-forming turbine on a 66~kV collector network with a $\pm$525~kV bipolar DRU link show that the commanded power is delivered over 800 to 1000~MW with the offshore voltage held near nominal. The paper also examines how grid-forming operation changes the allocation of control duties within the Type-4 turbine: once the grid-side converter forms the network, DC-link regulation must move to the machine-side converter, which limits the achievable bandwidth of the DC-link control. Extending the system to 2~GW with two 500~MW grid-following units achieves full power transfer in simulation, but with the grid-forming converter loaded beyond its continuous rating and a steady-state offshore frequency deviation of approximately 1.8~Hz.
\end{abstract}

\maketitle

\section{Introduction}
\label{sec:intro}

As offshore wind farms move farther from shore and grow in rating, high-voltage direct-current (HVDC) transmission becomes the preferred option for bringing the power ashore. The charging current of a submarine AC cable grows with its length and occupies part of its current rating, which limits the length of a high-voltage AC (HVAC) connection but not of an HVDC one \cite{korompili2016}. In the established arrangement, an offshore voltage-source converter forms the offshore AC network and is also the heaviest and most expensive item used in this topology. Diode-rectifier-unit (DRU)-HVDC replaces it with uncontrolled diode bridges, removing the offshore switching devices, their cooling, and much of the accompanying equipment \cite{blasco2010,yu2024review}.

The price of this simplification is a control problem. An uncontrolled rectifier can neither establish offshore voltage and frequency nor supply the reactive power that its own commutation consumes. These functions must come from the grid-side converters (GSCs) of the wind turbines, so grid-forming (GFM) operation is a physical requirement rather than a preferred mode \cite{blasco2010,yu2018pll,yu2024review}. This raises the question of how the turbines should do this: which GFM control strategy can both form the offshore network and regulate the power delivered through an uncontrolled rectifier, and which strategy cannot?

Several strategies have been proposed in answer: distributed voltage and frequency droop \cite{blasco2010}, modified virtual synchronous machine (VSM) schemes \cite{bidadfar2021}, power-synchronisation control \cite{karaca2025}, remote active-power regulation through the onshore terminal \cite{wang2024}, adaptive reactive-power allocation \cite{gzhang2025}, and hybrid plants mixing GFM and grid-following (GFL) turbines \cite{jin2025}. They differ mainly in the variable through which active power is controlled, either the converter angle, as in VSM and power-synchronisation control, or the voltage magnitude.

Existing results already point to the voltage magnitude, rather than the converter angle, as the variable that sets the power delivered through a
DRU. Power transfer through a diode rectifier is set by the difference between the AC and DC voltage magnitudes~\cite{blasco2010,wang2024}. From
the dynamics of the offshore network,~\cite{herrera2017} concluded that active power couples to voltage and reactive power to
frequency, the reverse of a conventional grid, an assignment adopted by several subsequent controllers~\cite{xiao2022,zhang2024}. Even where angle-based control has been used, as in the droop and VSM schemes that~\cite{beneit2025} operated together in one DRU-connected plant, the total power was set by a separate plant-level controller acting on the voltage magnitude, and the angle-based loops only shared that power among the turbines. What is missing is the other half of the argument. In~\cite{zhang2024}, researchers derive the sensitivities of the DRU equations that motivate the P-V and Q-$\omega$ assignment, but neither their work nor the other works show why the angle-based alternative fails. 
% Two questions therefore remain open: why does angle-based control fail on a DRU, so that an additional voltage-magnitude loop is needed, and what does the Q-$\omega$ loop of a GFM unit actually control?

Existing DRU GFM studies are also typically at or below 1~GW and include AC filters or reactive compensation at the rectifier, for example the 1~GW, $\pm$320~kV system of~\cite{yu2018pll}, the 400~MW monopolar system of~\cite{bidadfar2021} and the 200~MW system of~\cite{zhang2024}, and they commonly represent each turbine by its GSC fed from a constant DC source~\cite{yu2018pll,zhang2024}, while~\cite{herrera2017} uses grid-following turbines with an external frequency controller. Such models hide how the GFM duty interacts with the rest of the turbine and how much reactive power the converters must supply in the absence of filters, so whether a suitable controller scales to a full 2~GW plant remains open.

This paper addresses these gaps at industrial scale. Its contributions are: (i) the DRU commutation equation is used to show why VSM angle control fails on a DRU, since the delivered power has no direct control path through the converter angle, and hence why the P-V/Q-$\omega$ architecture of \cite{zhang2024} suits it; (ii) the architecture is implemented in PSCAD/EMTDC on a 1~GW aggregated Type-4 turbine, tuned by pole placement for the inner loops and electromagnetic transient (EMT) simulation for the outer loops, and validated across operating setpoints; (iii) the GFM turbine is modelled with both its machine-side converter (MSC) and GSC in a 2~GW, $\pm$525~kV bipolar system with two 24-pulse DRUs and no AC filters, which exposes the conflict between the MSC and GSC over DC-link regulation and places the full DRU reactive demand on the GFM converter; and (iv) the design is extended to a 2~GW hybrid GFM-GFL configuration, showing that the active power injected by the grid-following units raises the reactive power the GFM converter must supply, which with a fixed reactive reference shifts the offshore frequency.

\section{System description and main circuit design}
\label{sec:system}

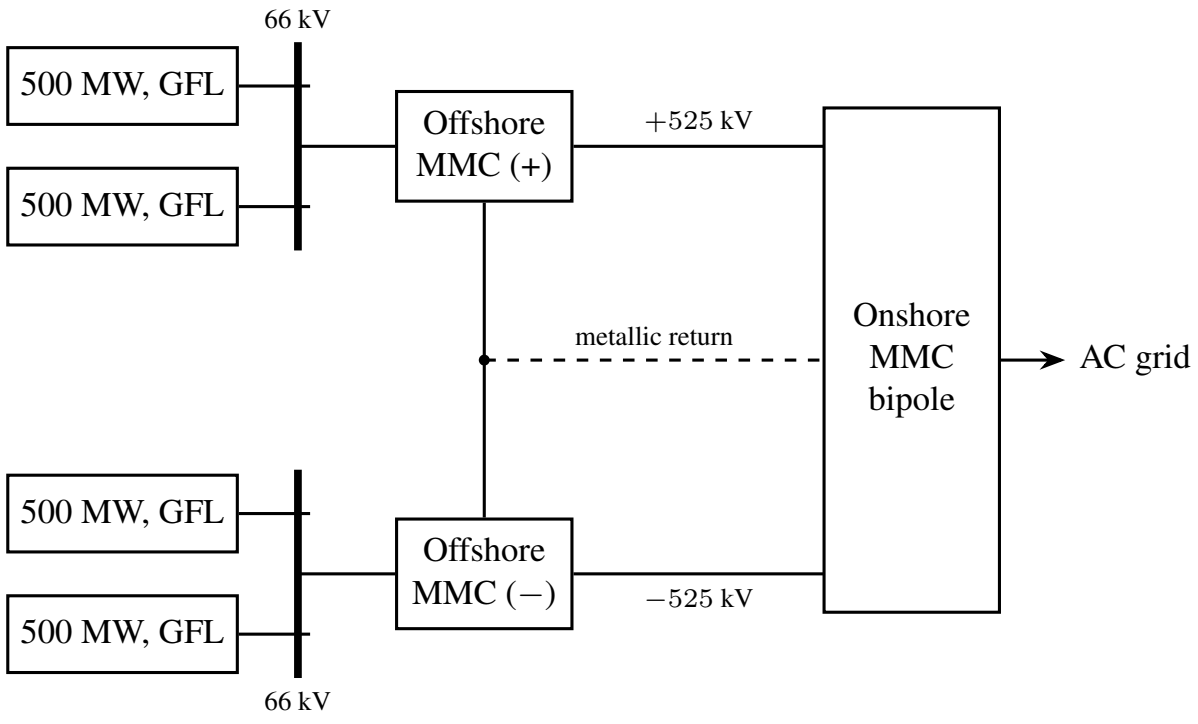
\begin{figure}[!t]
\centering
\resizebox{0.9\linewidth}{!}{%
\begin{tikzpicture}[
  wt/.style={draw, thick, minimum width=1.9cm, minimum height=0.7cm, align=center, font=\small},
  cnv/.style={draw, thick, minimum width=1.6cm, minimum height=1.0cm, align=center, font=\small},
  >=Stealth]
% turbines, positive pole
\node[wt] (w1) at (0, 2.5) {500~MW, GFL};
\node[wt] (w2) at (0, 1.4) {500~MW, GFL};
% turbines, negative pole
\node[wt] (w3) at (0,-1.4) {500~MW, GFL};
\node[wt] (w4) at (0,-2.5) {500~MW, GFL};
% 66 kV buses
\draw[line width=2pt] (1.6,1.0) -- (1.6,2.9);
\draw[line width=2pt] (1.6,-2.9) -- (1.6,-1.0);
\node[font=\scriptsize, above] at (1.6,2.9) {66~kV};
\node[font=\scriptsize, below] at (1.6,-2.9) {66~kV};
\foreach \w in {w1,w2,w3,w4} \draw[thick] (\w.east) -- ++(0.65,0);
% offshore MMC poles
\node[cnv] (mp) at (3.3, 1.95) {Offshore\\MMC (+)};
\node[cnv] (mn) at (3.3,-1.95) {Offshore\\MMC ($-$)};
\draw[thick] (1.6,1.95) -- (mp.west);
\draw[thick] (1.6,-1.95) -- (mn.west);
% midpoint and metallic return
\draw[thick] (mp.south) -- (3.3,0) -- (mn.north);
\fill (3.3,0) circle (0.05);
% DC cables
\draw[thick] (mp.east) -- (6.4,1.95) node[pos=0.5, above, font=\scriptsize] {$+525$~kV};
\draw[thick, dashed] (3.3,0) -- (6.4,0) node[pos=0.5, above, font=\scriptsize] {metallic return};
\draw[thick] (mn.east) -- (6.4,-1.95) node[pos=0.5, below, font=\scriptsize] {$-525$~kV};
% onshore MMC bipole
\node[draw, thick, minimum width=1.6cm, minimum height=4.6cm, align=center, font=\small] (on) at (7.2,0) {Onshore\\MMC\\bipole};
\draw[thick, ->] (on.east) -- (8.6,0) node[right, font=\small] {AC grid};
\end{tikzpicture}}
\caption{Original TenneT model: four aggregated 500~MW grid-following turbines, two feeding the positive-pole and two the negative-pole offshore MMC, with the $\pm$525~kV bipolar link and metallic return to the onshore MMC. In this study the offshore MMCs are replaced by two 24-pulse DRUs on a common collector bus, Fig.~\ref{fig:system}.}
\label{fig:original}
\end{figure}

The study starts from a 2~GW PSCAD/EMTDC model provided by TenneT, in which each pole of the $\pm$525~kV bipole is fed by a pole of an offshore modular multilevel converter (MMC) and four aggregated grid-following Type-4 turbines of 500~MW (560~MVA, 112 units of 5~MVA) are connected, two per pole, Fig.~\ref{fig:original}. The positive and negative poles of the offshore MMC were disconnected and replaced with two 24-pulse DRUs, and all turbines were connected to a common 66~kV collector bus that feeds both DRUs.

\begin{figure}[!t]
\centering
\resizebox{\linewidth}{!}{%
\begin{tikzpicture}[lbl/.style={font=\small, align=center}]
% wind turbine
\draw[thick] (0.15,-0.9) -- (0.25,1.5) -- (0.35,1.5) -- (0.45,-0.9) -- cycle;
\foreach \a in {90,210,330}
  \draw[thick, rotate around={\a:(0.3,1.6)}] (0.85,1.6) ellipse (0.55 and 0.09);
\draw[thick, fill=white] (0.3,1.6) circle (0.08);
% PMSG
\draw[thick] (1.6,0) circle (0.5);
\draw (1.6,0) circle (0.36);
\node[lbl] at (1.6,-0.85) {PMSG};
\draw[thick] (2.1,0) -- (2.9,0);
\foreach \x in {2.4,2.5,2.6} \draw (\x-0.05,-0.1) -- (\x+0.05,0.1);
% MSC
\draw[thick] (2.9,-0.6) rectangle (4.1,0.6);
\draw (2.9,-0.6) -- (4.1,0.6);
\node at (3.2,0.3) {$\sim$};
\node at (3.8,-0.3) {$=$};
\node[lbl] at (3.5,-0.85) {MSC};
% DC link
\draw[thick] (4.1,0.4) -- (5.3,0.4);
\draw[thick] (4.1,-0.4) -- (5.3,-0.4);
\draw (4.7,0.4) -- (4.7,0.07);
\draw[thick] (4.5,0.07) -- (4.9,0.07);
\draw[thick] (4.5,-0.07) -- (4.9,-0.07);
\draw (4.7,-0.07) -- (4.7,-0.4);
\node[font=\scriptsize, right] at (4.85,0) {$C_\mathrm{dc}$};
\node[font=\scriptsize, above] at (4.7,0.4) {$V_\mathrm{dc}$};
% GSC
\draw[thick] (5.3,-0.6) rectangle (6.5,0.6);
\draw (5.3,-0.6) -- (6.5,0.6);
\node at (5.6,0.3) {$=$};
\node at (6.2,-0.3) {$\sim$};
\node[lbl] at (5.9,-0.85) {GSC};
% LC filter: series Lf, shunt Cf
\draw[thick] (6.5,0) -- (6.8,0);
\draw[thick] (6.8,0) arc (180:0:0.1) arc (180:0:0.1) arc (180:0:0.1) arc (180:0:0.1);
\node[font=\small, above] at (7.2,0.15) {$L_f$};
\draw[thick] (7.6,0) -- (8.6,0);
\fill (8.0,0) circle (0.04);
\draw (8.0,0) -- (8.0,-0.45);
\draw[thick] (7.8,-0.45) -- (8.2,-0.45);
\draw[thick] (7.8,-0.6) -- (8.2,-0.6);
\draw (8.0,-0.6) -- (8.0,-0.95);
\draw (7.85,-0.95) -- (8.15,-0.95);
\draw (7.9,-1.03) -- (8.1,-1.03);
\draw (7.95,-1.11) -- (8.05,-1.11);
\node[font=\small, right] at (8.2,-0.52) {$C_f$};
% turbine transformer
\draw[thick] (9.0,0) circle (0.4);
\draw[thick] (9.5,0) circle (0.4);
\node[font=\scriptsize] at (9.25,-0.65) {Transformer};
% PCC bus and DRU
\draw[thick] (9.9,0) -- (10.6,0);
\draw[line width=2pt] (10.6,-0.7) -- (10.6,0.7);
\node[font=\scriptsize, above] at (10.6,0.7) {PCC, 66~kV};
\draw[thick, ->] (10.6,0) -- (11.5,0) node[right, font=\small] {DRU};
\end{tikzpicture}}
\caption{Aggregated Type-4 wind turbine: PMSG, MSC, DC link, GSC with LC filter ($L_f$, $C_f$) and turbine transformer to the 66~kV PCC.}
\label{fig:wt}
\end{figure}
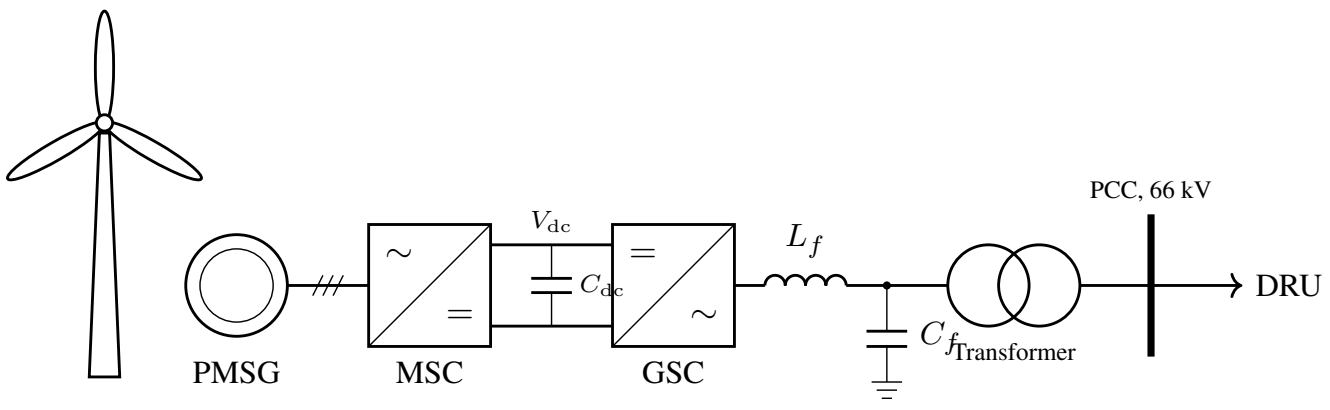

The final studied system, shown in Fig.~\ref{fig:system} of Section~\ref{sec:rating}, is a 2~GW offshore wind farm connected through a bipolar $\pm$525~kV DRU-HVDC link with metallic return to an onshore bipolar MMC link. Aggregated Type-4 turbines with permanent-magnet synchronous generators (PMSGs) feed a common 66~kV collector bus: WT1 (1000~MW, 200$\times$5~MVA, a rating justified in Section~\ref{sec:rating}) operates GFM, while WT2 and WT3 (500~MW, 560~MVA each) operate GFL. The point of common coupling (PCC) is the terminal of WT1 at the 66~kV collector bus, where $V_\mathrm{pcc}$ is measured. Each turbine has an MSC, a DC link, and a GSC with an LC output filter, as shown in Fig.~\ref{fig:wt}. Except where stated otherwise, each turbine is represented by its GSC, with the DC link modelled as an ideal source, thereby isolating the GFM control from the machine-side dynamics.

\begin{figure}[!t]
\centering
\includegraphics[width=0.9\linewidth]{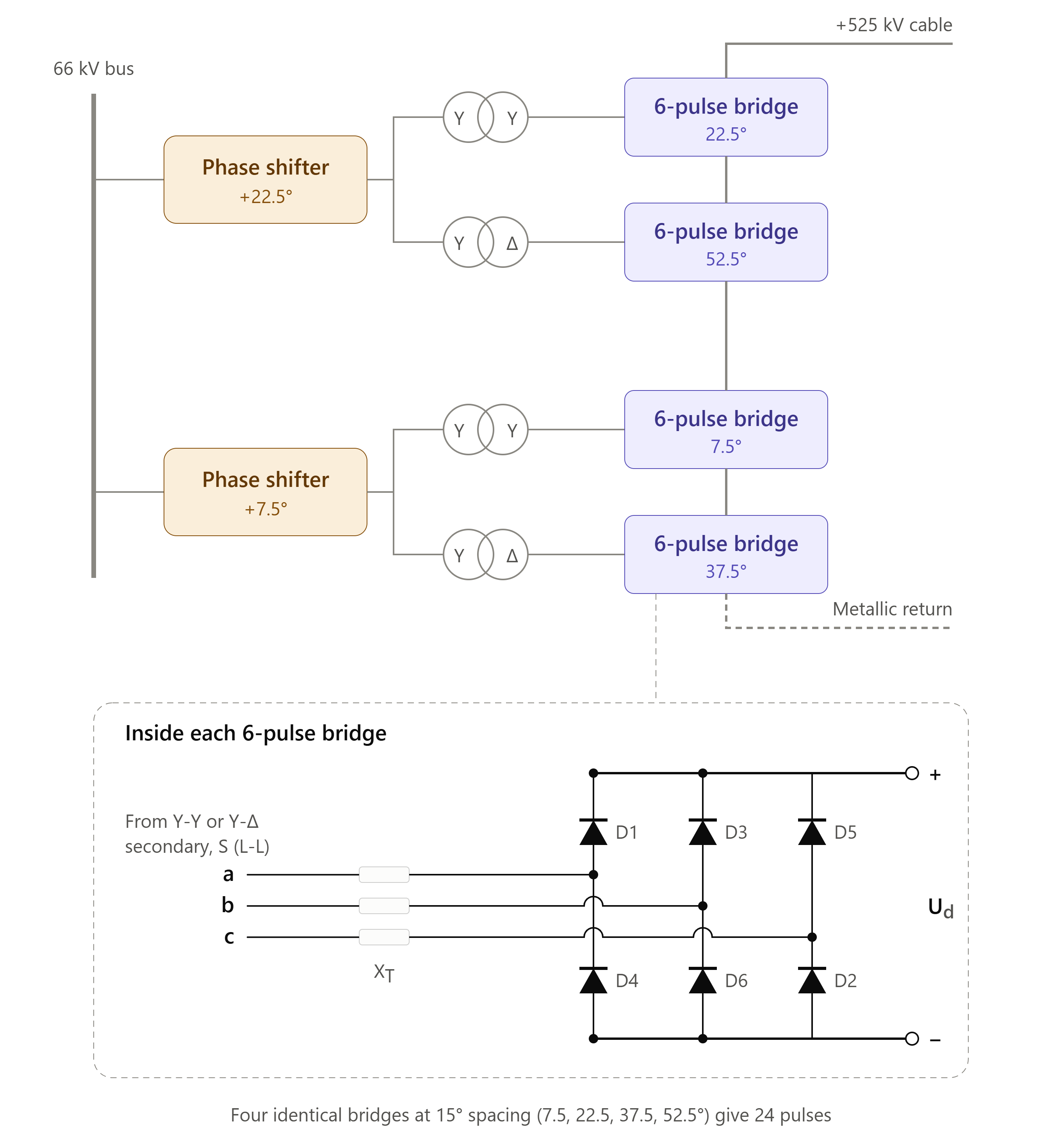}
\caption{One 24-pulse DRU pole: two phase-shift transformers, four bridge transformers and four series 6-pulse diode bridges.}
\label{fig:dru}
\end{figure}

Each pole is fed by a 24-pulse DRU of four series-connected 6-pulse bridges supplied through two phase-shift transformers ($+22.5^\circ$, $+7.5^\circ$) and Y-Y/Y-$\Delta$ bridge transformers, Fig.~\ref{fig:dru}, cancelling characteristic harmonics below the 23rd order~\cite{arrillaga1998}. The averaged DC pole voltage follows the commutation equation:
\begin{equation}
U_\mathrm{dru} = \frac{4\times 3\sqrt{2}}{\pi}\,\frac{U_0}{66}\,S - \frac{12}{\pi}\,X_T I_\mathrm{dc},
\label{eq:commutation}
\end{equation}
where $U_\mathrm{dru}=525$~kV, $U_0$ is the collector line-to-line voltage in kV, $S$ the bridge-transformer secondary voltage in kV, $I_\mathrm{dc}$ the DC current, and $X_T$ the commutation reactance per bridge, which aggregates all series reactances in the AC current path between the 66~kV collector bus and the diode valves; the $\pm$525~kV DC cable carries DC current only and plays no role in commutation. Two properties of (\ref{eq:commutation}) shape this work: it contains the voltage magnitude $U_0$ but no angle, and its second term reflects commutation overlap. The leakage reactance $X_T$ prevents the current from transferring instantly from one diode to the next, so two diodes conduct together for a short interval, known as the overlap angle. As a result, the AC current lags the voltage, and the DRU consumes reactive power that the diode bridge cannot supply and that must come from the turbine converters or from compensation on the platform.

The bridge-transformer ratio is 66/102~kV, chosen so that the DRU transmits the 1~GW design power of the first study stage, that is 500~MW per pole ($I_\mathrm{dc}\approx 952$~A), at rated collector voltage. With this ratio, the ideal no-load DC voltage of the pole, given by the first term of (\ref{eq:commutation}) at $U_0 = 66$~kV, is $U_\mathrm{d0} = 551$~kV; in operation the commutation drop brings the DRU output down to the 525~kV held by the onshore converter. The reactive power consumed by the commutation overlap is~\cite{zhang2024}
\begin{equation}
Q_\mathrm{DRU} = P\sqrt{\left(\frac{U_\mathrm{d0}}{U_\mathrm{dru}}\right)^{2} - 1}
= 0.319\,P ,
\label{eq:qdru}
\end{equation}
about 319~MVAr at 1~GW.
\section{Methodology}
\label{sec:method}

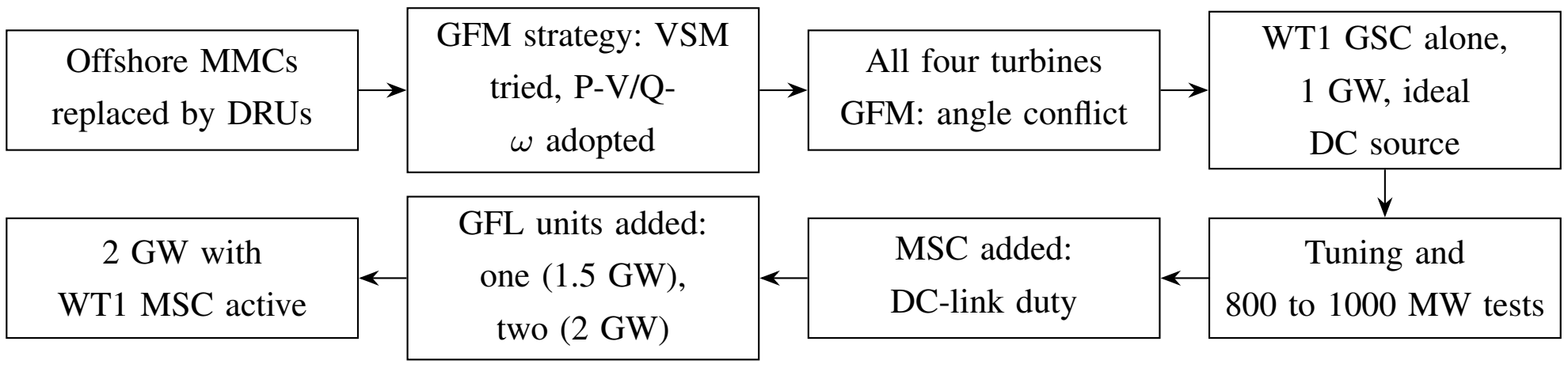
\begin{figure*}[t]\centering
\begin{tikzpicture}[node distance=4mm, box/.style={draw, align=center, font=\footnotesize, minimum height=10mm, text width=27mm}, >=Stealth]
\node[box] (a) {Offshore MMCs\\replaced by DRUs};
\node[box, right=of a] (b) {GFM strategy: VSM\\tried, P-V/Q-$\omega$ adopted};
\node[box, right=of b] (c) {All four turbines\\GFM: angle conflict};
\node[box, right=of c] (d) {WT1 GSC alone,\\1~GW, ideal DC source};
\node[box, below=of d] (e) {Tuning and\\800 to 1000~MW tests};
\node[box, left=of e] (f) {MSC added:\\DC-link duty};
\node[box, left=of f] (g) {GFL units added:\\one (1.5~GW), two (2~GW)};
\node[box, left=of g] (h) {2~GW with\\WT1 MSC active};
\draw[->] (a)--(b); \draw[->] (b)--(c); \draw[->] (c)--(d); \draw[->] (d)--(e); \draw[->] (e)--(f); \draw[->] (f)--(g); \draw[->] (g)--(h);
\end{tikzpicture}
\caption{Staged methodology: each stage isolates one problem before the next is added.}
\label{fig:stages}
\end{figure*}

The study followed an isolate-and-solve approach. Replacing the offshore MMC with a DRU changes several parts of the system at once: the turbine GSCs must form the offshore voltage and frequency, the active-power control path moves from the converter angle to the voltage magnitude, the DC-link duty inside the turbine moves to the MSC, and the full commutation reactive demand falls on the turbine converters. Because these changes interact, the study was developed in stages: each stage isolates one problem and solves it before the next is added, until the whole system works (Fig.~\ref{fig:stages}). After the offshore MMCs of the reference model were replaced by DRUs, the GFM strategy was selected by testing on the GSC the VSM and the P-V/Q-$\omega$ control described in Section~\ref{sec:control}. The controller was then applied to all four turbines, after which a single GFM turbine, represented by its GSC with an ideal DC source and enlarged from 500~MW to 1~GW (Section~\ref{sec:rating}), was tuned and tested from 800 to 1000~MW. The MSC was added next to take over the DC-link duty, followed by the grid-following units, first one (1.5~GW) and then two (2~GW), and finally the 2~GW case was run with the MSC active. Each stage, and the choice it led to, is reported in Section~\ref{sec:results}.

\section{Grid-forming control design}
\label{sec:control}

Since the wind turbines must operate grid-forming, the first question is which GFM control can regulate the power delivered through the DRU, and which cannot.

\subsection{VSM angle control and the DRU power path}

GFM control is most commonly realised by emulating a synchronous machine, and the VSM is one of its most established forms; it has also been applied to DRU-connected plants~\cite{bidadfar2021,beneit2025}. It is therefore the natural first candidate and is examined first.
% At the start of this work, before the constraints that the DRU places on the control variables were known, it was therefore the natural first choice for forming the offshore network.

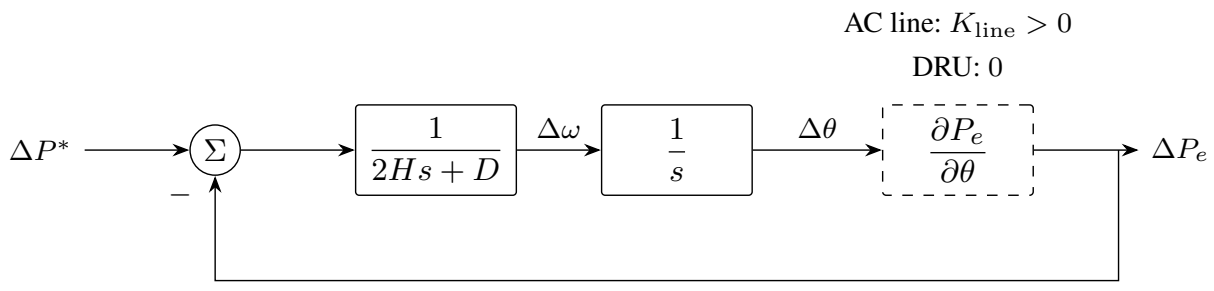
\begin{figure}[!t]
\centering
\resizebox{0.92\linewidth}{!}{%
\begin{tikzpicture}[
  blk/.style={draw, rounded corners=1pt, minimum height=0.9cm, minimum width=1.5cm, align=center, font=\small},
  sum/.style={draw, circle, minimum size=0.5cm, inner sep=0pt, font=\small},
  lbl/.style={font=\footnotesize},
  >=Stealth]
\node[sum] (s) at (0,0) {$\Sigma$};
\node[blk] (sw) at (2.2,0) {$\dfrac{1}{2Hs + D}$};
\node[blk] (int) at (4.6,0) {$\dfrac{1}{s}$};
\node[blk, dashed] (plant) at (7.4,0) {$\dfrac{\partial P_e}{\partial\theta}$};
\draw[->] (-1.3,0) node[left, lbl] {$\Delta P^{*}$} -- (s.west);
\draw[->] (s.east) -- (sw.west);
\draw[->] (sw.east) -- node[above, lbl] {$\Delta\omega$} (int.west);
\draw[->] (int.east) -- node[above, lbl] {$\Delta\theta$} (plant.west);
\draw[->] (plant.east) -- (9.2,0) node[right, lbl] {$\Delta P_e$};
\draw[->] (9.0,0) -- (9.0,-1.3) -- (0,-1.3) -- (s.south);
\node[lbl] at (-0.35,-0.45) {$-$};
\node[lbl, align=center] at (7.4,1.05) {AC line: $K_\mathrm{line}>0$\\ DRU: $0$};
\end{tikzpicture}}
\caption{VSM synchronising loop: the DRU provides no angle-to-total-power gain, so the active-power loop remains open.}
\label{fig:signalflow}
\end{figure}

A VSM forms voltage and frequency through the swing equation:
\begin{equation}
2H\frac{d\omega}{dt} = P^{*} - P_e - D(\omega-\omega_0), \qquad \frac{d\theta}{dt} = \omega,
\label{eq:swing}
\end{equation}
so the active-power error acts on the converter angle. Since the delivered power, $P_e = U_\mathrm{dru} I_\mathrm{dc}$ with $I_\mathrm{dc}$ from (\ref{eq:commutation}), contains no angle, linearising it gives
\begin{equation}
\Delta P_e = \underbrace{\frac{\partial P_e}{\partial U_0}}_{K_\mathrm{DRU}>0}\Delta U_0 +\underbrace{\frac{\partial P_e}{\partial\theta}}_{=\,0}\Delta\theta .
\label{eq:jacobian}
\end{equation}
Equation (\ref{eq:jacobian}) shows that the delivered power is not controllable through the angle: the gain from the VSM output $\theta$ to the controlled output $P_e$ is zero, so the active-power loop of (\ref{eq:swing}) remains open; see Fig.~\ref{fig:signalflow}. Angle-based loops may still share power among parallel converters, but the total power must be controlled through the voltage magnitude.

\subsection{Adopted P-V/Q-$\omega$ architecture}
With $U_\mathrm{dru}$ held by the onshore converter, (\ref{eq:commutation}) shows that the DC current, and hence the power delivered per pole, $P = U_\mathrm{dru} I_\mathrm{dc}$, rises with the collector voltage magnitude $U_0$~\cite{zhang2024}. Active power must therefore be regulated by the voltage magnitude, thereby forming a P-V loop. The angle then carries no active-power information, so the frequency needs a different reference, and the reactive power provides it. For the offshore capacitance $C$, formed by the GSC filter and cable capacitance, the q-axis equation in a frame rotating at $\omega$ is $C\,dv_q/dt = i_q - i_{q,\mathrm{dru}} - \omega C v_d$. In a frame aligned with the PCC voltage, $v_q = 0$ and $dv_q/dt = 0$, so this equation becomes algebraic~\cite{herrera2017},
\begin{equation}
\omega\, C\, V_d = i_q - i_{q,\mathrm{dru}},
\label{eq:qbalance}
\end{equation}
where $i_q$ is the q-axis current injected by the turbines and $i_{q,\mathrm{dru}}$ the q-axis current drawn towards the DRU. The frequency is therefore fixed by the reactive-current balance and the voltage by the active current~\cite{herrera2017}. Researchers in~\cite{zhang2024} reach the same assignment from the sensitivities of the DRU equations, $\partial P/\partial U > 0$ and $\partial Q/\partial f = -(Q_\mathrm{dru}+Q_\mathrm{fil})/f$, where $Q_\mathrm{fil}$ is the reactive power of the shunt capacitance, which gives the P-V and Q-$\omega$ loops.

% The P-V/Q-$\omega$ control of ~\cite{zhang2024} has exactly this structure and is adopted because it was designed for Type-4 GSCs on a DRU-HVDC link and provides the complete cascade, validated on a 12-pulse DRU system of about 200~MW, with the turbine DC links represented by ideal sources and with AC filter banks and reactive compensation at the rectifier.

The P-V/Q-$\omega$ control of~\cite{zhang2024} has exactly this structure and is adopted because it was designed for Type-4 GSCs on a DRU-HVDC link and provides the complete cascade. Here it is applied without AC filters, so the converter supplies the full DRU reactive demand. Its start-up and fault-ride-through functions are not used. Active power acts on the voltage magnitude,
\begin{equation}
E_\mathrm{ref} = E_0 + \left(K_{p,\mathrm{PV}} + \frac{K_{i,\mathrm{PV}}}{s}\right)(P^{*}-P),
\label{eq:pv}
\end{equation}
where $E_\mathrm{ref}$ is the d-axis voltage reference passed to the GSC voltage loop (the q-axis reference is zero), $E_0 = 1$~pu its nominal value, $P^{*}$ and $P$ the reference and measured active power of WT1, and $K_{p,\mathrm{PV}}$, $K_{i,\mathrm{PV}}$ the PI gains. This is the closed path identified in (\ref{eq:jacobian}): raising the GSC voltage raises the PCC voltage, the DC current and the transmitted power. The P-V loop settles at whatever voltage the DRU needs to transmit $P^{*}$; the 66/102~kV ratio of Section~\ref{sec:system} places this voltage at 1.0~pu for the 1~GW design point, so that the converter keeps its voltage headroom. The frequency is generated from the reactive-power error,
\begin{equation}
\omega = \omega_0 - \frac{K_G}{1+sK_T}\,(Q - Q_\mathrm{ref}), \qquad \theta = \int \omega\, dt,
\label{eq:qw}
\end{equation}
so the GSC is its own frequency source, without a phase-locked loop, and the Q-$\omega$ loop also aligns the $dq$ frame. Here $\omega_0$ is the nominal angular frequency, $Q$ the reactive power delivered by WT1, and the frequency falls when it exceeds $Q_\mathrm{ref}$; this sign differs from the form written in~\cite{zhang2024}. The reference is set slightly below the commutation demand of (\ref{eq:qdru}), at $Q_\mathrm{ref} = Q^{*} = 0.30P$.

\begin{figure*}[!t]
\centering
\includegraphics[width=0.65\textwidth]{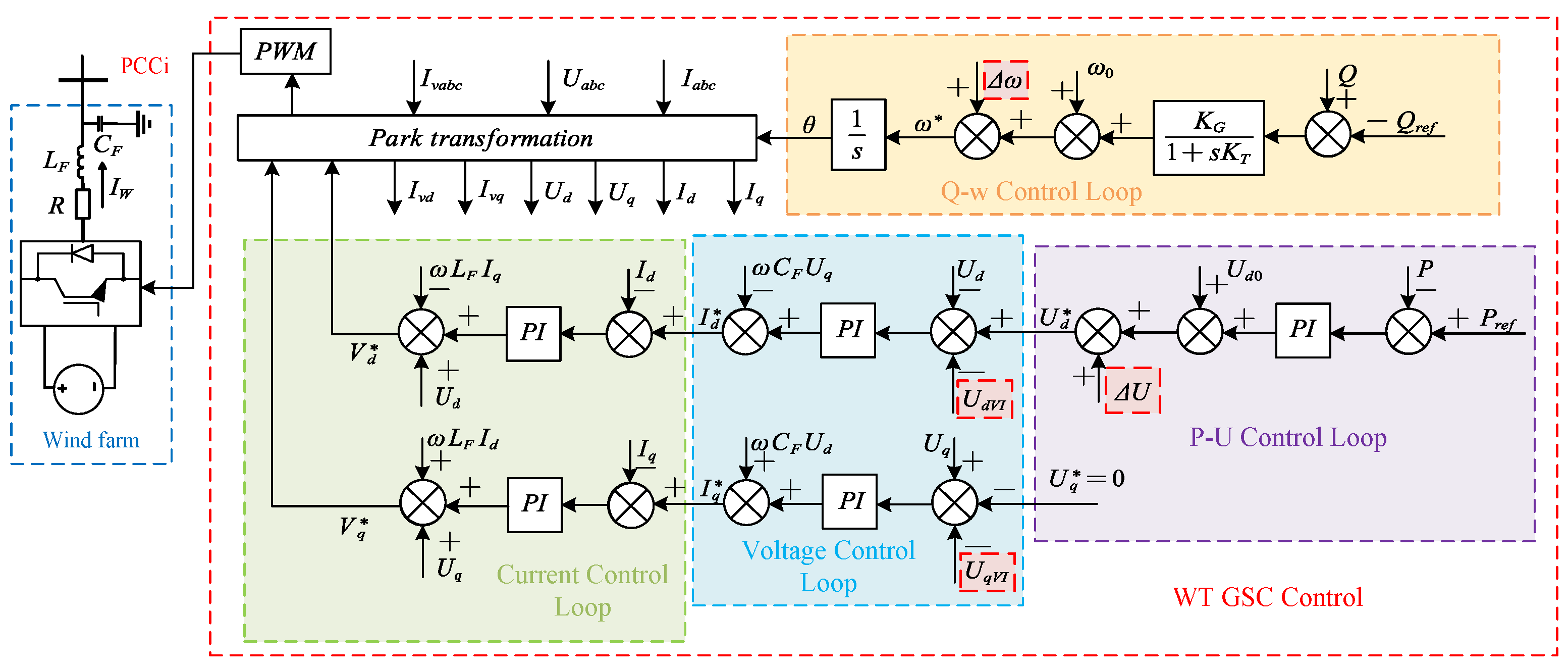}
\caption{GSC GFM control structure, reproduced from \cite{zhang2024}. The P-V loop corresponds to (\ref{eq:pv}); the $\Delta U$ and $\Delta\omega$ inputs pertain to start-up and fault-ride-through functions that are not used here, and the virtual impedance is fixed.}
\label{fig:cascade}
\end{figure*}

The complete cascade, Fig.~\ref{fig:cascade}, comprises a $dq$ current loop with cross-coupling feedforward, a voltage loop with virtual impedance for LC-filter damping \cite{dahono2001}, the P-V loop, and the Q-$\omega$ loop. Zhang et al.~\cite{zhang2024} adapt the virtual impedance during low-voltage ride-through to limit fault current; since fault ride-through is outside the scope of this work, a fixed virtual impedance is used.

\subsection{Tuning procedure}
\label{sec:tuning}

The cascade of Fig.~\ref{fig:cascade} must be tuned before it can be implemented; it is tuned for the 1~GW GFM unit used in the rest of the paper. In this work tuning was the most time-consuming step, and after sequential loop-by-loop tuning proved unsatisfactory, the gains were set by bandwidth separation between the nested loops and pole placement of the inner loops, in three steps (see Table~\ref{tab:bandwidth}). First, the current-loop bandwidth is set to $\omega_\mathrm{CC}=\omega_0/5$ and the voltage loop is placed a factor of five lower \cite{yazdani2010,teodorescu2011}. Second, the current and voltage loops have integrating plants $G(s)=1/(s\tau)$, with $\tau$ the per-unit filter inductance or capacitance, and the gains of the PI controller $K_p(1 + 1/(sT_i))$, with proportional gain $K_p$ and integral time constant $T_i$, are chosen by pole placement~\cite{yazdani2010},
\begin{equation}
K_p = \tau\,\omega_\mathrm{bw}, \qquad T_i = 1/\omega_\mathrm{bw},
\label{eq:imc}
\end{equation}
which places the closed-loop poles at the natural frequency $\omega_\mathrm{bw}$ with a damping ratio of 0.5. With the filter values of Fig.~\ref{fig:wt}, $L_f = 0.106$~mH and $C_f = 2.65$~mF per 5~MVA, 600~V unit, $\tau = 0.463$ and $0.060$~pu on the turbine base, giving $K_p = 29.09$ and $0.754$ with $T_i = 0.016$ and $0.0796$~s for the current and voltage loops. Third, because the outer loops act on the nonlinear DRU relation, the P-V gains are set in EMT simulation. For the Q-$\omega$ loop, the gain $K_G = 0.114$ was initialised from the reactive-power coefficient of~\cite{zhang2024} and, like the time constant $K_T = 0.4$~s, retained after trial in EMT simulation. All outer-loop gains are verified over the 800 to 1000~MW operating range.

\begin{table}[!t]
\caption{Cascade bandwidths and tuning basis}
\begin{center}
\begin{tabular}{lll}\toprule
Loop & Bandwidth or gains & Tuning \\\midrule
Current    & $\omega_0/5 = 62.8$~rad/s & pole placement, (\ref{eq:imc}) \\
Voltage    & $\omega_\mathrm{CC}/5 = 12.6$~rad/s & pole placement, (\ref{eq:imc}) \\
P-V        & set in EMT & EMT \\
Q-$\omega$ & $K_G = 0.114$, $K_T = 0.4$~s & from \cite{zhang2024}, EMT \\\botrule
\end{tabular}
\label{tab:bandwidth}
\end{center}\vspace*{-12pt}
\end{table}

\section{Results and discussion}
\label{sec:results}

\subsection{Control strategy and configuration}

%A VSM was implemented on the GSC first, but the delivered power could not be regulated through it, although the DRU, the MSC, and the GSC inner loops each behaved as expected when tested in isolation. This is consistent with the open loop predicted by (\ref{eq:jacobian}): the angle has no path to the delivered power. The P-V/Q-$\omega$ control of Section~\ref{sec:control} was therefore used in all further stages.

Implemented on the GSC, the VSM could not regulate the delivered power, as predicted by (\ref{eq:jacobian}), so the P-V/Q-$\omega$ control was used in all further stages. With the P-V/Q-$\omega$ controller applied to all four aggregated 500~MW turbines, each converter integrated its own angle from its own Q-$\omega$ loop, and the competing integrators produced persistent oscillations in power and frequency. Parallel GFM operation in DRU-connected plants requires additional coordination, such as secondary control, consensus or adaptive reactive allocation~\cite{blasco2010,sun2024,gzhang2025,beneit2025}. Here, the problem was isolated instead: a single turbine, WT1, was made GFM and represented by its GSC with an ideal DC source, while the other turbines were disconnected and later added as GFL units (Section~\ref{sec:hybrid}), consistent with~\cite{jin2025}.

\subsection{Rating of the grid-forming unit}
\label{sec:rating}

WT1 is rated 1~GW (200 units of 5~MVA), replacing two of the original 500~MW turbines. Until the grid-following units are connected, the GFM turbine carries the DRU alone, so its rating sets the margin above the conduction threshold of the DRU. The onshore converter holds the HVDC cable at $U_\mathrm{dru} = 525$~kV regardless of the offshore power, so the DRU diodes conduct only when the rectified collector voltage exceeds the cable voltage~\cite{herrera2017,zhang2024}. Since $U_\mathrm{d0} = 551\,U_0$~kV with $U_0$ in pu (Section~\ref{sec:system}), this conduction threshold is $U_0 = 525/551 \approx 0.953$~pu. Above it, rearranging (\ref{eq:commutation}), the collector voltage $U_0(P)$, in pu, needed to transmit a total power $P$ over both poles is
\begin{equation}
U_0(P) = \frac{U_\mathrm{dru} + \Delta U_\mathrm{c}(P)}{551~\mathrm{kV}}, \qquad \Delta U_\mathrm{c}(P) = \Delta U_\mathrm{c,1GW}\,\frac{P}{1~\mathrm{GW}},
\label{eq:u0p}
\end{equation}
where $\Delta U_\mathrm{c}(P) = 12 X_T I_\mathrm{dc}/\pi$ is the commutation drop. Since it is proportional to $I_\mathrm{dc}$ and hence to $P$, it can be scaled from its value $\Delta U_\mathrm{c,1GW}$ at 1~GW. At 1~GW the PCC voltage is about 1.0~pu (Section~\ref{sec:validation}), so $\Delta U_\mathrm{c,1GW} \approx 551 - 525 = 26$~kV. The same relation predicts $U_0 \approx (525 + 52)/551 = 1.047$~pu at 2~GW, close to the 1.04~pu observed in Section~\ref{sec:hybrid}. The delivered power therefore changes by about $K_\mathrm{DRU} = 551/26 \approx 21$~pu per pu of collector voltage.

\begin{figure}[!t]
\centering
\includegraphics[width=0.9\linewidth]{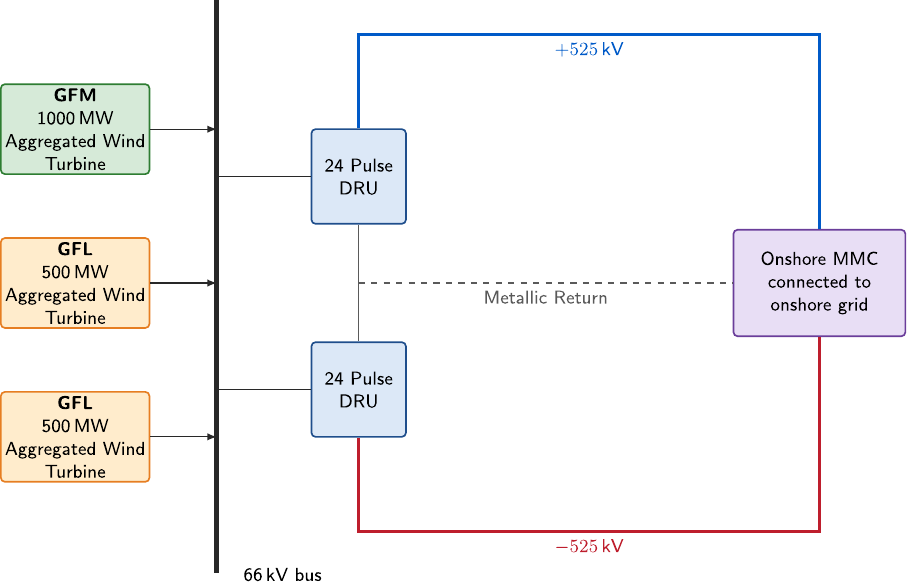}
\caption{Topology of the 2~GW DRU-HVDC offshore wind system: one 1000~MW GFM and two 500~MW GFL aggregated turbines, 66~kV collector bus, two 24-pulse DRUs and a bipolar $\pm$525~kV link.}
\label{fig:system}
\end{figure}

From (\ref{eq:u0p}), a 500~MW unit would operate at $U_0 = (525 + 13)/551 \approx 0.976$~pu, only 0.024~pu above the threshold, against 0.047~pu at 1~GW. With a plant this steep, a transient voltage dip of that size stops conduction altogether. Consistent with this, a 500~MW case, with the gains of Section~\ref{sec:tuning} tuned around 1~GW, showed a sustained power oscillation of about 1.4~Hz in which the PCC voltage reached the threshold in every cycle, whereas at 800 to 1000~MW the corresponding transient decays. In addition, as the only GFM unit, a 500~MW WT1 would give a GFM share of 25\,\% at 2~GW, below the critical share of about 37\,\% reported in~\cite{jin2025}, against 50\,\% at 1~GW. The minimum rating of the GFM unit was not determined because the aim of the study is to establish whether the DRU can carry the full 2~GW; the aggregated unit was instead enlarged to 1~GW, which doubles the margin above the threshold. The resulting configuration is shown in Fig.~\ref{fig:system}.

\subsection{Single grid-forming turbine validation}
\label{sec:validation}

WT1 alone forms the offshore network at 800, 900 and 1000~MW; the steady states are listed in Table~\ref{tab:steadystate}.

\begin{table}[!t]
\caption{Single GFM turbine steady state ($Q^{*}=0.30P$)}
\begin{center}
\begin{tabular}{lcccc}\toprule
$P$ (MW) & $Q$ (MVAr) & $V_\mathrm{pcc}$ (pu) & $f$ (Hz) & $Q/P$ \\\midrule
800  & 220.0 & $\approx$1.0 & 50.020 & 0.275 \\
900  & 259.5 & $\approx$1.0 & 49.896 & 0.288 \\
1000 & 300.6 & $\approx$1.0 & 49.755 & 0.301 \\\botrule
\end{tabular}
\label{tab:steadystate}
\end{center}\vspace*{-12pt}
\end{table}

The commanded power is delivered at all setpoints with the PCC voltage within 1\,\% of 1.0~pu, confirming that the 66/102~kV ratio places the natural operating point at the design target, and the frequency remains within 0.25~Hz of 50~Hz. The measured 300.6~MVAr at 1000~MW is within 6\,\% of the commutation estimate (\ref{eq:qdru}), and the rise of $Q/P$ from 0.275 to 0.301 is consistent with the overlap angle growing with DC current. The reactive surplus $Q - Q_\mathrm{ref}$ grows from $-20$~MVAr at 800~MW to about zero at 1000~MW, while the frequency falls from 50.02 to 49.76~Hz. The deviation does not follow the mismatch in proportion, so in the single-turbine cases the frequency is not set by the Q-$\omega$ droop alone, and the reference is an operating-point choice rather than a universal $Q/P$ constant.

\subsection{Machine-side converter integration}
\label{sec:msc}

Replacing the ideal DC source by the MSC of WT1 raises the question of which converter regulates the DC link. Its voltage obeys $C_\mathrm{dc}\,dV_\mathrm{dc}/dt=P_\mathrm{MSC}-P_\mathrm{GSC}$, so one of the two converters must keep the two powers balanced. In a grid-following turbine the roles follow from the hardware. The MSC is connected directly to the generator and is the only converter that can set its torque, so it controls the rotor speed for maximum-power-point tracking, while the GSC regulates $V_\mathrm{dc}$ by exporting whatever power the MSC delivers into a grid formed elsewhere. In GFM operation the GSC outer loops are occupied by the P-V/Q-$\omega$ control, which itself sets the exported power, so the GSC can no longer balance the DC link and the duty must pass to the MSC, the reverse of the conventional assignment. This is a conflict rather than a simple swap. Because the MSC acts through the generator torque, the rotor and drivetrain now sit inside the DC-link loop: every change in the power drawn by the GSC must be met by a change in generator torque, and the MSC can no longer track the optimum rotor speed independently, so maximum-power-point tracking would have to enter through the power reference $P^{*}$ of the GSC instead.

\begin{figure}[!t]
\centering
\includegraphics[width=0.8\linewidth]{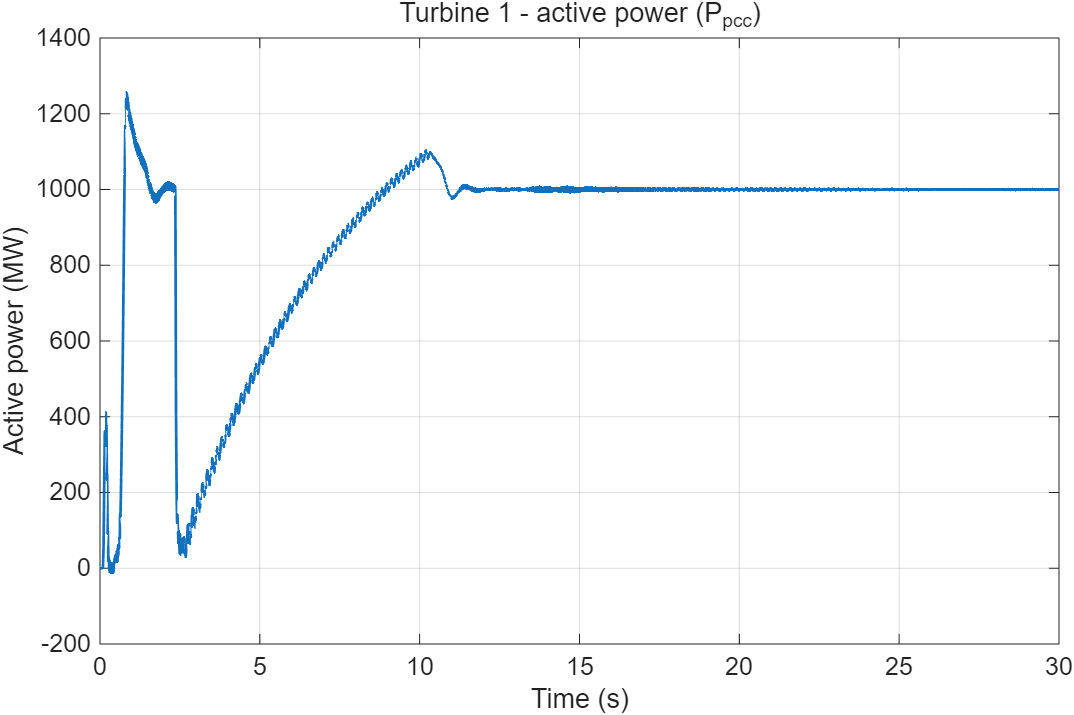}
\caption{Active power of WT1 at the PCC during start-up with the MSC active. The auxiliary DC source is disconnected and the MSC deblocked at about 2.3~s.}
\label{fig:msc}
\end{figure}

With the DC-link duty assigned to the MSC, the GSC is deblocked first while an auxiliary DC source holds the link at 1.6~kV. At about 2.3~s the source is disconnected and the MSC is deblocked, so its PI loop takes over the link in a single step. The active power then drops almost to zero, Fig.~\ref{fig:msc}, and the DC-link voltage to about 0.5~kV. With the MSC DC-link loop deliberately made slow, the system recovers only gradually: the power returns to its setpoint after about 8~s, with an overshoot to about 1100~MW near 10~s, and the DC-link voltage settles back at 1.6~kV only after about 20~s. A faster loop, with an integral time of the order of 0.02~s, excited a growing oscillation in $V_\mathrm{dc}$ within a few seconds of deblocking, ending in integrator windup and collapse, consistent with DC-link/drivetrain interaction reported for GFM PMSG turbines \cite{liu2024}.

\subsection{Hybrid GFM-GFL extension to 2~GW}
\label{sec:hybrid}

The grid-following units are then added. All hybrid cases use the transformer ratio and controller gains of the 1~GW design, and in each WT1 first forms the network and settles alone. With one grid-following unit (1.5~GW), WT2 is then deblocked at 500~MW and 150~MVAr; with two (1.6 to 2~GW), WT2 and WT3 are deblocked together at the same setpoints.

\begin{table}[!t]
\caption{Hybrid configurations (WT2, WT3: GFL, 500~MW)}
\begin{center}
\begin{tabular}{lcccl}\toprule
WT1 (GFM) & Total & GFM (\%) & $f$ (Hz) & Outcome \\\midrule
1000~MW$^{a}$ & 1.5~GW & 67 & $\approx$49.0 & stable \\
1000~MW       & 2.0~GW & 50 & $\approx$48.2 & stable \\
800~MW        & 1.8~GW & 44 & $\approx$48.7 & stable \\
600~MW        & 1.6~GW & 37.5 & n/a & unsettled \\\botrule
\end{tabular}\\[2pt]
{\footnotesize $^{a}$WT3 not connected.}
\label{tab:hybrid}
\end{center}\vspace*{-12pt}
\end{table}

\begin{figure*}[!t]
\centering
\begin{minipage}[t]{0.485\textwidth}\centering
\includegraphics[width=\linewidth]{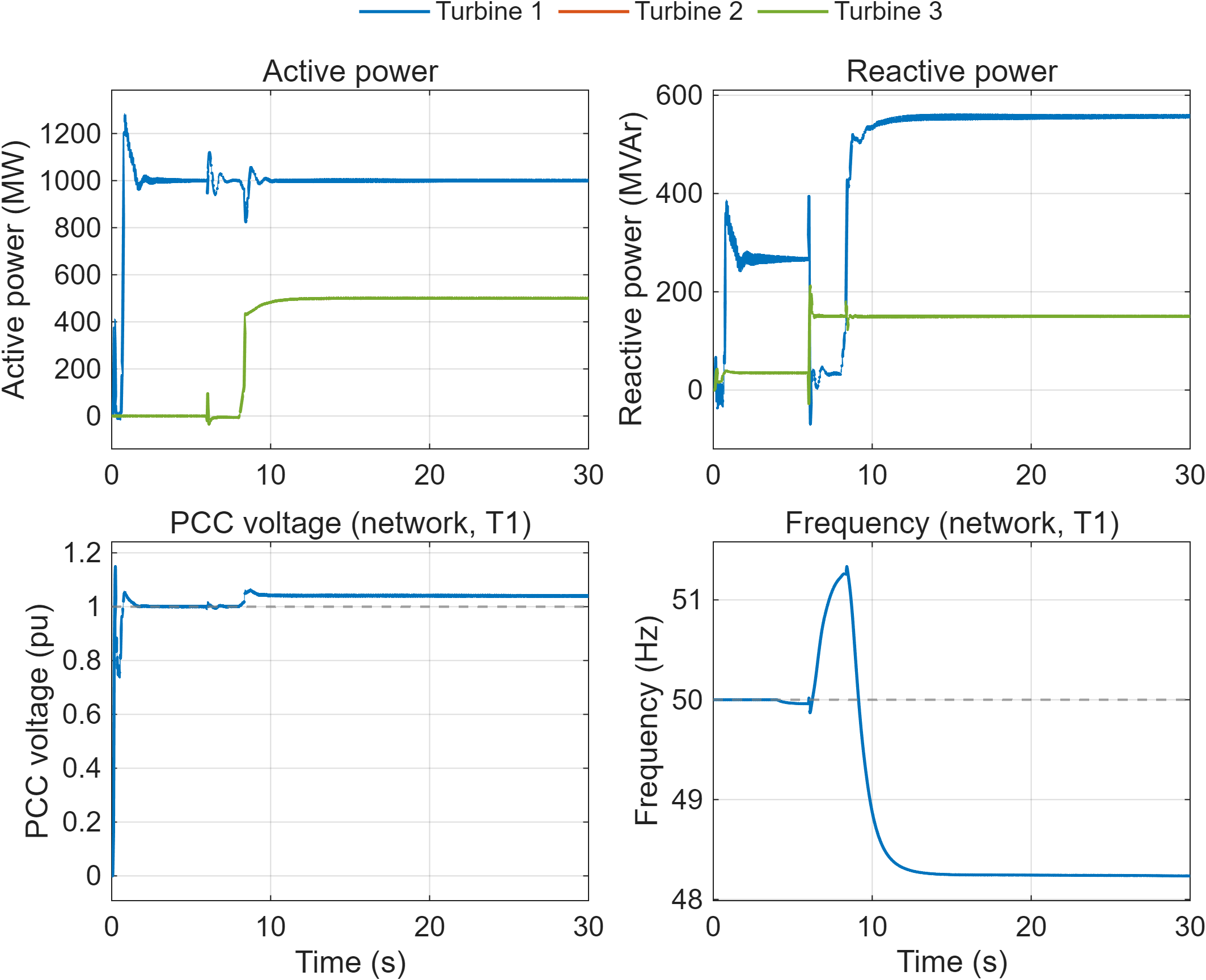}
\caption{EMT response of the 2~GW hybrid configuration: WT1 GFM at 1000~MW, WT2 and WT3 GFL at 500~MW each (WT2 and WT3 curves coincide).}
\label{fig:2gw}
\end{minipage}\hfill
\begin{minipage}[t]{0.485\textwidth}\centering
\includegraphics[width=\linewidth]{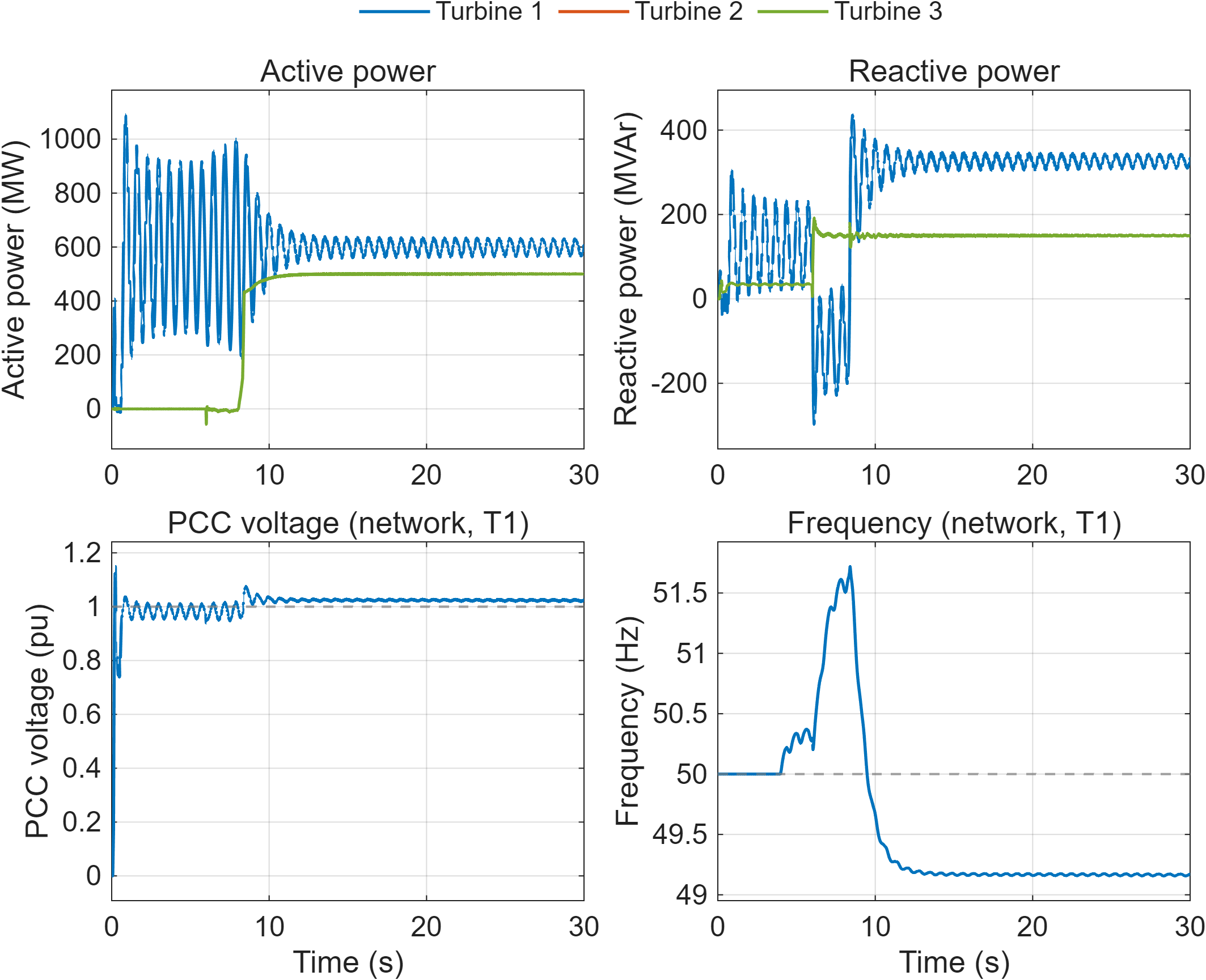}
\caption{Reduced GFM share: WT1 GFM at 600~MW with WT2 and WT3 GFL at 500~MW each (1.6~GW); the system does not settle.}
\label{fig:reduced}
\end{minipage}
\end{figure*}

Table~\ref{tab:hybrid} summarises the configurations. The first three reached and held their target power in the EMT runs, as confirmed by measurements through the DRU, with the frequency settling to a bounded deviation; Fig.~\ref{fig:2gw} shows the full 2~GW case. With WT1's MSC active (Section~\ref{sec:msc}) and WT1 settled before WT2 and WT3 are deblocked, the 2~GW case again reaches 1000/500/500~MW, a PCC voltage of about 1.04~pu and a frequency of about 48.2~Hz, the same as with the ideal source. The 37.5\% GFM-share case exhibited a sustained oscillation as seen in Fig.~\ref{fig:reduced}. This is consistent with the reduced-GFM-share limitation reported in \cite{jin2025}, but the present study does not establish a minimum GFM share, since reduced GFM capacity and operating-point-dependent tuning were not separated.

The central finding of the extension is the coupling between GFL injection, the GFM reactive operating point, and the offshore frequency. WT1's reactive output rises from about 300~MVAr alone to 425 to 430~MVAr with WT2 and about 550~MVAr with WT2 and WT3. Each GFL unit's injection raises the power through the DRU and, with it, the commutation reactive demand. At 2~GW, (\ref{eq:u0p}) gives $U_0 \approx 1.047$~pu, so $U_\mathrm{d0} \approx 577$~kV and (\ref{eq:qdru}) gives $Q_\mathrm{DRU} \approx 0.46 \times 2000 \approx 910$~MVAr, close to the 850~MVAr supplied by WT1, WT2 and WT3 together. WT1's reference, however, remains at the 1~GW design value, and with the sign of (\ref{eq:qw}) the growing surplus lowers the frequency to about 49.0~Hz with one GFL unit and 48.2~Hz with two. The DC link isolates the offshore frequency from the onshore grid, so this violates no synchronisation requirement; it is nonetheless undesirable, as it shifts the cascade bandwidths and may excite network-filter resonances.

\subsection{Discussion and limitations}

Unlike the current-source turbines of~\cite{herrera2017}, whose injected reactive current sets the frequency through (\ref{eq:qbalance}), a GFM GSC imposes the frequency and the network draws the reactive power it needs. Since $\partial Q/\partial f$ is small, about 6~MVAr per Hz at 1~GW, the Q-$\omega$ loop appears to set the frequency from the reactive mismatch rather than regulate $Q$; its role is to fix the frequency and $dq$ frame and, with several GFM units, to share reactive power through their angles~\cite{zhang2024}.

The DRU reactive demand must be supplied offshore. At 2~GW, WT1 delivers 1000~MW and about 550~MVAr, an apparent power of about 1140~MVA against its 1000~MVA nameplate; already at 1~GW standalone it reaches about 1045~MVA. This is within the converter current limit but beyond its continuous rating. The burden can be shared with the GFL units, which here supply 150~MVAr each out of about 250~MVAr of headroom at 560~MVA, or carried by platform compensation such as AC filters or a STATCOM; \cite{jin2025} shows that the minimum GFM share is set by this current loading and falls when filters carry part of the demand. Because the Q-$\omega$ loop of a GFM unit cannot move its own reactive output, removing the frequency deviation requires changing $Q_\mathrm{ref}$ or supplying the demand elsewhere: for example, an outer loop that adjusts $Q_\mathrm{ref}$ until the frequency returns to 50~Hz, as in the central PI controller of~\cite{herrera2017}, of which the fixed $Q^{*}=0.30P$ used here is only the feedforward part, or secondary frequency restoration~\cite{blasco2010,sun2024}, adaptive reactive-power allocation~\cite{gzhang2025} and GFM operating-point compensation~\cite{jin2025}.

The 2~GW result rests on a single 1000~MW aggregated GFM converter. The bridge-transformer ratio was chosen at the 1~GW design point; the commutation overlap and transformer loading at 2~GW, where the DC current doubles, the DC overvoltage towards $U_\mathrm{d0}$ after a loss of the onshore converter, the intra-group interaction of individually controlled turbines and the margin of $E_\mathrm{ref}$ to its limits were not examined.

\section{Conclusion}
\label{sec:conclusion}

DRU-HVDC removes the conventional source of offshore voltage and frequency regulation, so grid formation must be provided by the wind turbines. Because the DRU commutation relationship contains the offshore voltage magnitude but no converter angle, the delivered power has no control path through the angle. A VSM implemented on the GSC could not regulate the power, and the P-V/Q-$\omega$ architecture of~\cite{zhang2024}, which regulates active power through the voltage magnitude, was adopted instead. Implemented on a 1~GW aggregated GSC with a 66/102~kV bridge-transformer ratio chosen for the 1~GW design point, it delivered 800 to 1000~MW with the PCC voltage near 1.0~pu. With the MSC included, DC-link regulation moves from the GSC to the MSC, and raising the bandwidth of this loop led to growing oscillation, so it had to be kept slow; as a result, the turbine needs about 20~s to recover after the DC-link handover. Extending the system to 2~GW with two grid-following units achieved full power transfer in the model, but the active power of the grid-following units raised the reactive power WT1 must supply from about 300 to 550~MVAr, beyond its continuous rating, and with the fixed reactive reference the offshore frequency settled at about 48.2~Hz. These results indicate that grid-forming operation on a DRU affects the whole turbine and the plant-level reactive balance, not only the GSC control. Future work will address the MSC-GSC allocation problem, the minimum stable GFM share, multi-GFM coordination, reactive-power sharing and frequency restoration, and black start.

\section{Acknowledgements}

This work was carried out in collaboration with TenneT TSO B.V. within the framework of an MSc thesis project at Delft University of Technology. The work is also supported by Horizon Europe project PROSECCO, under grant agreement 101160687.

\section{References}
\bibliographystyle{unsrt}
\bibliography{GFMreferences}

\end{document}